\documentclass[conference]{IEEEtran}
\IEEEoverridecommandlockouts
\usepackage{cite,comment}
\usepackage{amsmath,amssymb,amsfonts}
\usepackage{algorithm}
\usepackage{algpseudocode}
\usepackage{graphicx}
\usepackage{textcomp}
\usepackage{xcolor}
\usepackage{braket}
\usepackage{float}
\usepackage{booktabs}
\usepackage{subcaption}
\usepackage{balance}
\usepackage{lipsum}
\usepackage{xcolor}

\usepackage{stfloats}
\usepackage{supertabular}

\allowdisplaybreaks 

\def\BibTeX{{\rm B\kern-.05em{\sc i\kern-.025em b}\kern-.08em
    T\kern-.1667em\lower.7ex\hbox{E}\kern-.125emX}}
\begin{document}

 
\title{Towards Efficient Synthesis of Quantum Graph States by Fusing Graph Motifs \\
}
\author{
\IEEEauthorblockN{Tingxiang Ji$^{\dag}$, Hansika Weerasena$^{\dag}$, Demitry Farfurnik$^{\star}$,
        Jianqing Liu$^{\dag}$
}
\IEEEauthorblockA {
  $^{\dag}$Department of Computer Science, North Carolina State University, Raleigh, USA 27606. \\
  $^{\star}$Department of Electrical and Computer Engineering, Department of Physics and Astronomy,\\ North Carolina State University, Raleigh, USA 27606.
}
\IEEEauthorblockA {
\{tji2, hlokuka, dfarfur, jliu96\}@ncsu.edu
}

}

\maketitle

\begin{abstract}

Photonic graph states with advanced topologies can enable measurement-based quantum computing, distributed quantum sensing, and quantum interconnects. However, the efficient generation of photonic graph states is limited by the probabilistic nature of photonic entangling operations and the exponential dependence of generation rate on resource cost. In this work, we study photonic graph state synthesis as a cost-aware decomposition problem, exploiting local Clifford (LC) equivalence to identify more synthesis-friendly representations of the target graph state before decomposition. Specifically, we propose Cost-aware Fusion-based Decomposition (CFD), a three-stage heuristic framework that decomposes a target graph state into ring, star, and linear motifs, and assembles them via Type-I fusion operations to minimize fusion overhead and physical-qubit consumption. We further show that selecting the LC-equivalent graph state with the minimum number of edges provides a highly effective proxy for near-optimal synthesis: in many cases it matches the best generation rate observed within the LC equivalence class under CFD, and in most remaining cases it remains close to it. Numerical evaluations on graph state orbit data and 2D and 3D lattice graph states demonstrate that CFD achieves up to $84.6\%$ reduction in resource overhead compared to baseline constructions, and yields improvements in photonic generation rate spanning multiple orders of magnitude. These results suggest that combining structure-aware motif decomposition with LC equivalence is a practical and scalable strategy for photonic graph state synthesis.
\end{abstract}


\section{Introduction}\label{introduction}



Large entangled graph states are multipartite quantum states in which many qubits are mutually entangled according to the connectivity pattern of a graph. Such states are a fundamental resource for measurement-based quantum computation (MBQC) and quantum networking~\cite{liu2025road}, where computation and communication tasks are driven by local operations and measurements on a shared entangled substrate \cite{hein2004multiparty, raussendorf2003measurement}. 

In photonic platforms, graph states are especially attractive for two key reasons: the ease of long-distance distribution enabled by photons as natural carriers of quantum information, and hardware compatibility. Deterministic two-qubit gates are notoriously difficult to implement in photonics, making the MBQC paradigm, which replaces gate operations with large entangled states and measurements, a far more natural fit than the traditional circuit model. This makes photonic MBQC a particularly compelling framework. However, generating large graph states with the size and connectivity required by real applications has become an important practical challenge.


A standard approach in photonic MBQC is to assemble a large target graph state from smaller entangled resource states using linear-optical fusion operations \cite{browne2005resource}. This paradigm has enabled a broad line of work on scalable graph-state growth, including fusion-based constructions and percolation-inspired architectures \cite{pant2019percolation, morley2018physical, gimeno2015three}. However, each Type-I fusion gate succeeds with probability $1/2$, and each failed attempt requires regeneration of the involved resource states \cite{gu2024fendi}. For a graph state requiring $k$ fusion operations, the effective generation rate scales as $(1/2)^k$, decreasing exponentially with the number of fusions. Each additional fusion step also introduces further photon loss and resource consumption, making the direct edge-by-edge construction of complex graph topologies experimentally prohibitive.

At the same time, recent progress in emitter-based photonic entanglement generation has made it increasingly feasible to produce certain structured graph states directly \cite{lindner2009proposal, cogan2023deterministic, istrati2020sequential, de2024spin, rempe2024fusion}. Sequential-emission platforms naturally support families such as short linear clusters, GHZ/star states, and related small graph-state primitives, while advances in cavity-enhanced emitters continue to improve brightness and repetition rate \cite{ding2016demand, schwartz2016deterministic, thomas2022efficient}. 
Yet many application-driven target graphs have irregular and complex connectivity, creating a mismatch between the graph states that hardware can produce efficiently and those that a given application may require. Bridging this gap calls for a principled synthesis framework that maps target graph states onto hardware-friendly building blocks, one that also exploits the inherent flexibility in how a graph state can be represented.

A key observation from graph state theory is that a graph state is not uniquely tied to a single graph representation, but is defined only up to local Clifford (LC) equivalence, where two graph states are LC-equivalent if and only if they are related by a sequence of local complementations on their underlying graphs \cite{van2004graphical, van2004efficient, hein2004multiparty}. This equivalence has important practical consequences for photonic synthesis: two LC-equivalent graph states represent the same underlying quantum resource, yet one may be substantially easier to generate than the other. For example, any graph state whose underlying graph is a complete graph can always be transformed into a GHZ-type motif, where we use the term motif to refer to small graph states that serve as primitive building blocks for synthesis. This particular motif has a star as its underlying graph and can be directly realized by many photonic hardware platforms, reducing the number of fusion operations required during assembly. Exploiting LC equivalence to minimize fusion count is therefore a natural strategy, since the graph state generation rate degrades exponentially with the number of fusions. However, prior works either focus on fusion-based growth with fixed resource states, or on percolation-based results tailored to specific target topologies; neither jointly optimizes over LC-equivalent graph state representations and hardware constraints for arbitrary target graph states.



Motivated by these observations, we study photonic graph state generation as a cost-aware, hardware-aware synthesis problem. Rather than constructing a target graph state directly, we propose a heuristic decomposition framework, which we term Cost-aware Fusion-based Decomposition (CFD), that first identifies an LC-equivalent representative with reduced synthesis complexity, then decomposes it into motifs that can be assembled into the LC-equivalent graph state with fewer fusion operations. Our main contributions are as follows.

\begin{itemize}
\item We formulate photonic graph state generation as a cost-aware decomposition problem over LC-equivalent graph representations, where the objective is to reduce fusion overhead under hardware realizable resource states.
\item We propose CFD, a three-stage graph decomposition framework that maps a target graph state to ring, star/GHZ, and linear motifs for fusion-efficient synthesis.
\item Using graph state orbit data and lattice benchmarks, we show that the LC-equivalent graph state with the minimum number of edges is a strong practical proxy for synthesis-friendly forms, and that CFD reduces resource overhead by up to $84.6\%$ and improves photonic generation rate over existing baselines.
\end{itemize}




The remainder of this paper is organized as follows. Section~\ref{preliminaries} reviews the required background on graph states, local complementation, resource states, and generation rate modeling, and surveys related work on graph-state synthesis. Section~\ref{method} formulates the cost-aware synthesis problem and presents our decomposition-based framework for photonic graph-state generation. Section~\ref{evaluation} provides experimental results on decomposition cost, baseline comparison, and photonic generation rate. Section~\ref{conclusion} concludes the paper.

\section{Preliminaries}\label{preliminaries}
\subsection{Graph States}

A graph state is associated with a simple undirected graph $G=(V,E)$, where each vertex $v \in V$ represents a qubit, and each edge $(u,v) \in E$ represents a controlled-$Z$ (CZ) entangling operation between the corresponding qubits. 

The graph state $|G\rangle$ is prepared by initializing all qubits to $|+\rangle$ states and applying CZ gates along all edges in $E$:

\[
|G\rangle = \prod_{(u,v)\in E} CZ_{u,v} \; |+\rangle^{\otimes |V|}.
\]

Throughout this paper, we focus on the structural properties of $G$, and treat the entanglement structure of $|G\rangle$ as fully characterized by the connectivity pattern of $G$. 

\begin{figure}[htbp]
    \centering
    \begin{subfigure}[b]{0.32\columnwidth}
        \centering
        \includegraphics[width=\linewidth]{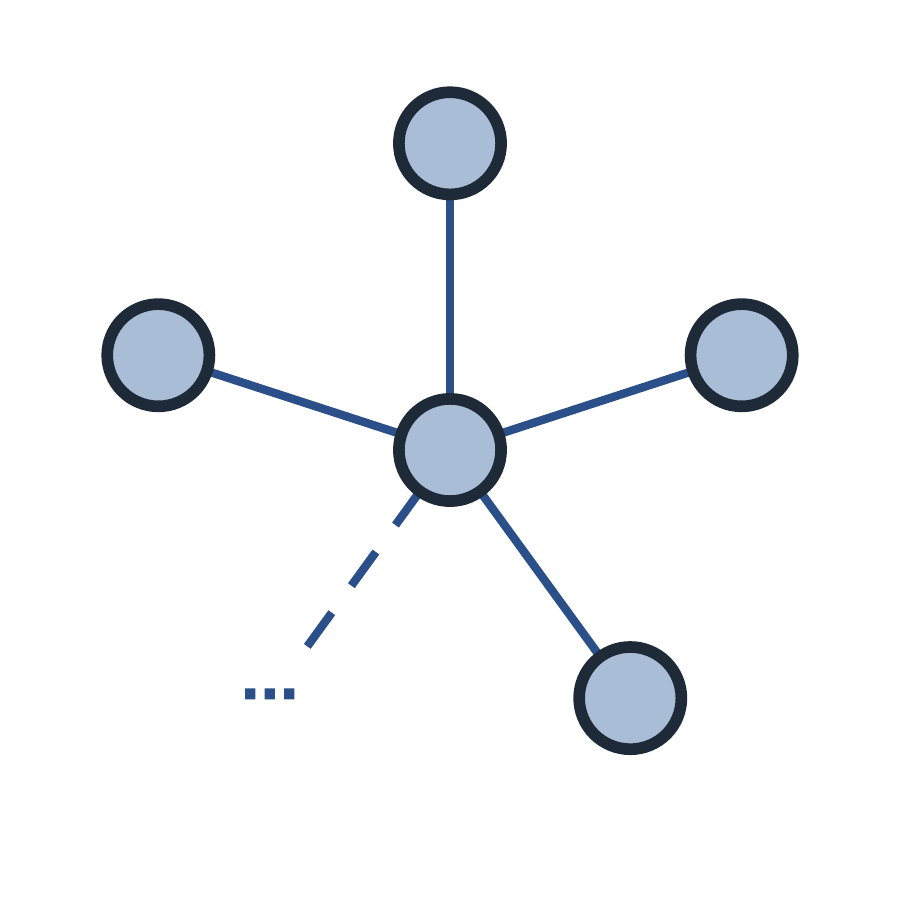}
        \caption{star motif}
        \label{fig:gs4}
    \end{subfigure}
    \hfill
    \begin{subfigure}[b]{0.32\columnwidth}
        \centering
        \includegraphics[width=\linewidth]{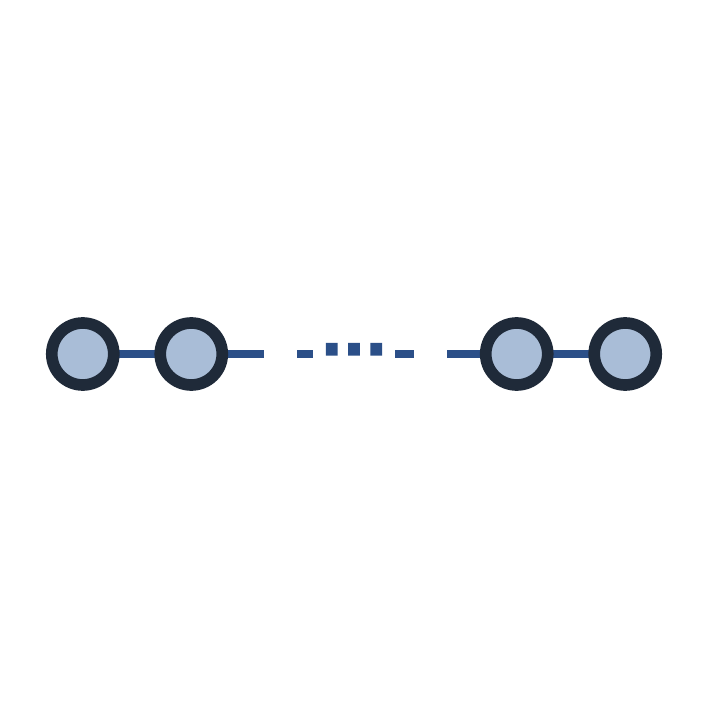}
        \caption{linear motif}
        \label{fig:star6}
    \end{subfigure}
    \hfill
    \begin{subfigure}[b]{0.32\columnwidth}
        \centering
        \includegraphics[width=\linewidth]{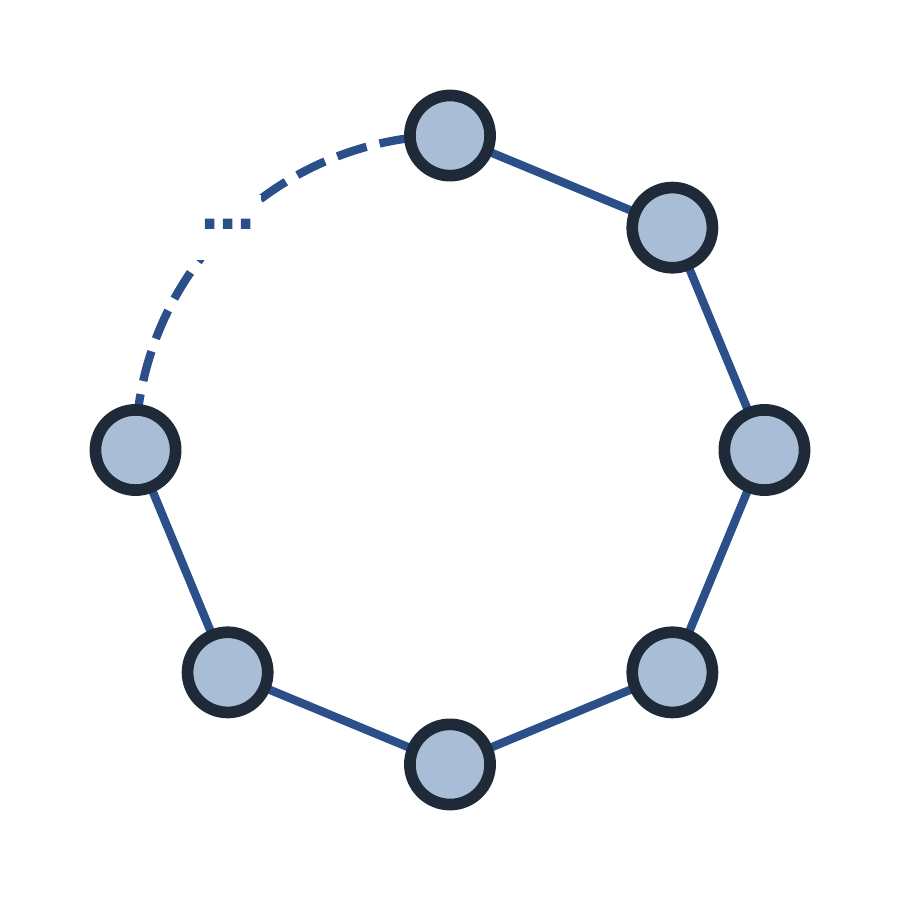}
        \caption{ring motif}
        \label{fig:gs4}
    \end{subfigure}
    \hfill
    \caption{Resource graph state motifs used in this work. (a) Star motif: a GHZ-type graph state with one central qubit connected to $N-1$ others, totaling $N$ qubits. (b) Linear motif: a path-like graph state with $N$ qubits connected sequentially. (c) Ring motif: a cyclic graph state with $N$ qubits. These three motifs serve as the primitive building blocks in the proposed CFD framework.}
    \label{fig:graph_states_all}
\end{figure}

\subsection{Resource States}

In this work, we consider a photonic platform in which a graph state is synthesized by fusing experimentally realizable resource states.
The resource states considered in this work include graph states with a star topology (referred to as \emph{star motif}), as shown in Fig.~\ref{fig:graph_states_all}(a); graph states with a path-like topology (referred to as \emph{linear motif}), as shown in Fig.~\ref{fig:graph_states_all}(b); and graph states with a ring topology (referred to as \emph{ring motif}), as shown in Fig.~\ref{fig:graph_states_all}(c). Throughout this paper, we use the term motif to refer to both the graph structure and its corresponding graph state: at the algorithmic level, a motif denotes a subgraph used in the decomposition; at the physical level, it refers to the resource graph state with that underlying topology.

Experimentally, the most straightforward resource states to generate are the star motif (GHZ state) Fig.~\ref{fig:graph_states_all}(a) and the linear motif Fig.~\ref{fig:graph_states_all}(b). The generation of such states typically involves the realization of the Lindner-Rudolph protocol on deterministic single photon emitters (cold atoms, trapped ions, or self-assembled quantum dots) that feature two ground spin states and two optically excited states~\cite{lindner2009proposal, cogan2023deterministic, schwartz2016deterministic, thomas2022efficient}. 
Based on the protocol, generating a star motif requires a repeated pulsed excitation of the spin and leveraging of optical selection rules, i.e., the emitter will decay to different spin states based on the polarization of the photon. Meanwhile, generating a linear motif requires a similar repeated excitation with the addition of $\pi/2$ pulses on the spin between excitations. Experimentally, the realization of $\pi/2$-pulses on the spin is possible by applying an optical or microwave pulse of coherent control or by applying a small external magnetic field and waiting for a quarter spin precession. The last type of resource state, namely the ring motif Fig.~\ref{fig:graph_states_all}(c) with $N$ photons, can be generated by fusing the first photon and the last photon of an $N+1$-qubit linear graph state by interfering these photons on a beamsplitter.

\subsection{Graph State Transformations}
We consider two classes of graph state transformations: local Clifford operations and fusion operations.

\paragraph{LC Operations}
LC operations are single-qubit Clifford gates applied independently to individual qubits. At the graph level, the fundamental LC operation is local complementation at a vertex $v$, denoted $\tau_v$, which transforms a graph $G=(V,E)$ into a new graph $\tau_v(G)$ by complementing the subgraph induced by the neighborhood $N(v)$: for every pair $\{u, w\} \subseteq N(v)$, the edge $\{u, w\}$ is added if it is absent and removed if it is present, while all other edges remain unchanged. Formally,
\[
E(\tau_v(G)) = E(G) \;\triangle\; \{\{u,w\} : u, w \in N(v),\; u \neq w\},
\]
where $\triangle$ denotes the symmetric difference.

Two graph states $\ket{G}$ and $\ket{G'}$ are LC-equivalent if one can be obtained from the other by a sequence of local complementations. LC-equivalent graph states represent the same entanglement resource up to local single-qubit operations, and can therefore be used interchangeably in synthesis tasks. 

\paragraph{Fusion Operations}

\begin{figure}
    \centering
    \includegraphics[width=0.8\linewidth]{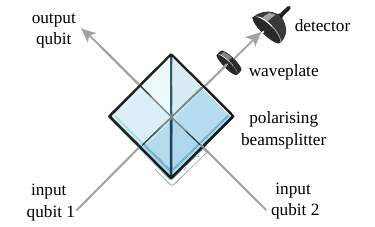}
    \caption{Schematic of the Type-I fusion in a photonic implementation. Two input qubits, one from each graph state, are interfered on a polarising beamsplitter followed by waveplate and detector, probabilistically merging the two fragments into a larger graph state with success probability $1/2$.}
    \label{fig:TypeI_Fusion}
\end{figure}

\begin{figure*}
    \centering
    \includegraphics[width=0.95\linewidth]{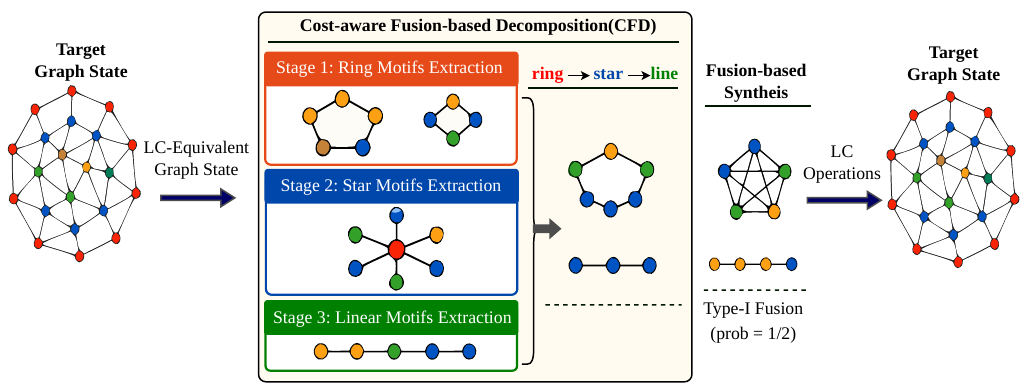}
    \caption{Overview of the proposed Cost-aware Fusion-based Decomposition (CFD) framework. Given a target graph state, CFD first identifies the LC-equivalent graph state with the minimum number of edges, then decomposes it into ring, star, and linear motifs through three successive extraction stages. The resulting motifs are realized as photonic resource states and connected via Type-I fusion operations to synthesize the LC-equivalent graph state, from which the target graph state is recovered by applying LC operations in the reverse order.}
    \label{fig:framework}
\end{figure*}

Fusion operations establish entanglement between separate graph state fragments by interfering photons on a beamsplitter followed by polarization-resolved detection \cite{browne2005resource, bartolucci2023fusion}. We focus on \emph{Type-I fusion}, which operates on one qubit from each of two graph states and succeeds with probability $1/2$. Fig.~\ref{fig:TypeI_Fusion} illustrates the Type-I fusion in a photonic implementation, where two input qubits are interfered on a polarising beamsplitter followed by waveplate and polarization-resolved detection.

At the graph level, consider two graph components $G_A$ and $G_B$ with designated fusion qubits $u \in V(G_A)$ and $v \in V(G_B)$. Intuitively, a successful Type-I fusion merges the two graph states by removing $v$ and redirecting all its connections to $u$: it removes $v$ while retaining $u$, and connects $u$ to every neighbor of $v$ in $G_B$. Formally, the resulting edge set is
\[
    E' = E(G_A) 
       \cup \bigl(E(G_B) \setminus \delta(v)\bigr) 
       \cup \bigl(\{u\} \times N(v)\bigr),
\]
where $\delta(v)$ denotes the set of edges incident to $v$, $N(v)$ denotes the neighbors of $v$ in $G_B$, and $\{u\} \times N(v)$ denotes the set of new edges connecting $u$ to every neighbor of $v$. The qubit $u$ retains all of its original edges in $G_A$. 

Upon failure, $v$ is measured out in the $Z$-basis and removed together with its incident edges, while $u$ remains intact in $G_A$. The two components therefore remain disconnected, but no damage is done to either fragment beyond the loss of $v$.

\subsection{Generation Rate Model}

The efficiency of photonic graph state generation is fundamentally constrained by photon loss and the probabilistic nature of fusion operations.

Let $\eta$ denote the overall photon collection efficiency, and $R_1$ the single-photon generation rate. The rate of generating an $N$-photon graph state without fusion is approximately
\[
R \approx \frac{R_1 \, \eta^N}{N},
\]
which decreases exponentially with the number of photons due to loss.

Fusion operations introduce additional overhead. In particular, a Type-I fusion gate succeeds with probability $1/2$ and consumes one photon from each participating fragment per attempt. For a graph state requiring $k$ successful fusion operations, each success must be obtained probabilistically, and failed attempts require regeneration of the involved resource states. As a result, the effective generation rate is 
\begin{equation}
R(G) \approx R_1 \cdot \frac{\eta^{N+k}}{N+k} \cdot \left(\frac{1}{2}\right)^k
\label{eq:generation_rate}
\end{equation}
where the additional factor $\eta^{k}$ captures the effective photon overhead introduced by fusion operations \cite{browne2005resource, bartolucci2023fusion}.

Reducing the number of fusion operations decreases photon loss and lowers the effective scaling exponent, yielding improvements across a broad range of hardware parameters.


\subsection{Background and Related Works}

Prior work on graph state synthesis has largely focused on two directions. The first line of work considers fusion-based growth strategies, where resource states are connected sequentially or in parallel to build up larger structures \cite{browne2005resource, bartolucci2023fusion}. Notably, FBQC \cite{bartolucci2023fusion} demonstrates how large-scale fault-tolerant quantum computation can emerge from fusion measurements on streams of small, fixed resource states. While powerful, these approaches typically assume a fixed set of resource states and do not exploit LC-equivalent reformulations of the target graph state to simplify synthesis.
The second line of work leverages percolation and graph-theoretic arguments to show that graph states with certain target topologies can be efficiently generated under probabilistic fusion \cite{pant2019percolation, morley2018physical}. However, these results are often tailored to specific target topologies or lattice geometries, and do not directly address the problem of synthesizing arbitrary target graph states from heterogeneous hardware primitives.
In contrast, our focus is topology-aware synthesis for arbitrary target graph states: we exploit LC-equivalent reformulations of the target graph state to identify more synthesis-friendly representations, decompose them into motifs, and assemble these motifs via fusion to reduce physical overhead on photonic platforms. To our knowledge, no existing framework jointly optimizes over LC-equivalent graph representations and hardware-aligned motif decompositions to minimize fusion overhead for arbitrary target graphs.

\section{Methodology}\label{method}

In this section, we first formulate the photonic graph state synthesis problem and then present the proposed decomposition framework.

\subsection{Problem Formulation}

Let $G = (V, E)$ be a simple undirected graph associated with a photonic graph state, as defined in Sec.~\ref{preliminaries}, where each vertex represents a photonic qubit and each edge corresponds to an entangling operation (e.g., a controlled-$Z$ gate).

In photonic architectures, large-scale graph states are typically not generated by directly implementing entangling operations for every edge in $E$. Instead, current experimental platforms naturally produce a restricted set of small entangled resource states, such as short linear cluster chains, star-shaped (GHZ-type) states, and small cyclic structures. Larger graph states are then constructed by connecting these resource states via probabilistic entangling operations, such as fusion measurements. This motivates a \emph{resource-synthesis} perspective: rather than directly realizing $G$, one seeks to assemble it from experimentally accessible building blocks.

Formally, let $G = (V, E)$ be the target graph state with $N = |V|$ photons. We seek a collection of motifs 
\[
\mathcal{S} = \{S_1, S_2, \dots, S_m\},
\]
that forms an edge-disjoint cover of $G$, where each $S_i$ belongs to a predefined family of experimentally realizable resource states, including ring motifs, star motifs, and linear motifs. Specifically, this requires
\[
E = \bigcup_{i=1}^{m} E(S_i), \qquad E(S_i) \cap E(S_j) = \varnothing \quad \text{for } i \neq j,
\]
so that each entangling edge is assigned to exactly one motif, while vertices may be shared across multiple motifs.

The motifs are connected via Type-I fusion operations. For a decomposition requiring $k$ fusion operations in total, the effective generation rate of the target graph state is given by Eq. \ref{eq:generation_rate}. Since both $\eta^k$ and $(1/2)^k$ decrease exponentially with $k$, the generation rate degrades rapidly as the number of fusion operations increases.


The synthesis task is therefore formulated as
\begin{equation}
\max_{\mathcal{S}} \; R_1 \cdot \frac{\eta^{N+k(\mathcal{S})}}{N + k(\mathcal{S})} \cdot \left(\frac{1}{2}\right)^{k(\mathcal{S})},
\label{eq:max_generation_rate}
\end{equation}
subject to $\mathcal{S}$ forming an edge-disjoint cover of $G$, where $k(\mathcal{S})$ denotes the total number of fusion operations required by the motif decomposition $\mathcal{S}$. Since $N$ is fixed for a given target graph and both penalty terms are monotonically decreasing in $k$, this is equivalent to minimizing the number of fusion operations:
\begin{equation}
\min_{\mathcal{S}} \; k(\mathcal{S}),
\label{eq:min_motifs}
\end{equation}
subject to the same edge-covering constraints. This reduced objective is the primary criterion used throughout our decomposition framework.

Finding an optimal edge-disjoint cover is combinatorially intractable in general, as the fragment selection decisions are interdependent and the search space grows exponentially with the number of edges. We therefore adopt a heuristic decomposition approach, described in the following section.

\subsection{CFD Framework}
An overview of the proposed framework is illustrated in Fig.~\ref{fig:framework}. In particular, our approach prioritizes hardware-friendly structures, namely the ring and star motifs, and covers the remaining edges using low-complexity linear motifs.

Our framework proceeds in three steps: we first identify an LC-equivalent representation of the target graph state with fewer edges, then apply the three-stage CFD to decompose it into hardware-friendly motifs by extracting ring, star, and linear motifs at the graph level, and finally connect the resulting motifs through Type-I fusion operations to realize the LC-equivalent graph state, from which the target graph state is recovered by applying LC operations in the reverse order.

We choose an LC-equivalent representation with fewer edges because a smaller number of edges corresponds to a simpler topology, reducing the number of required fusion operations. This is particularly important because the resource cost of fusion operations grows exponentially with fusion count, so even a modest reduction leads to significant savings in practice. Since LC operations are single-qubit gates which can be implemented by linear optics with near-unity fidelity and introduce negligible loss, recovering the target graph state from its LC-equivalent graph state adds minimal overhead to the overall synthesis process. We therefore neglect the cost of LC operations in our resource overhead analysis.


In practice, we obtain this LC-equivalent graph state as follows. We directly obtain the LC-equivalent graph state with the minimum number of edges by looking it up in a precompiled graph state dataset, which enumerates all graph states for $n=4$ to $9$ nodes. For graphs with $n \geq 10$ nodes, the method proposed in \cite{ji2024distributing} can be used to find an LC-equivalent graph state with fewer edges. The problem of finding such LC-equivalent representations is not the focus of this work.


The three motifs considered in this work, namely ring, star, and linear motifs, are directly realizable on current photonic hardware, as discussed in Section \ref{preliminaries}.
Specifically, CFD first extracts ring motifs, then identifies star motifs, and finally organizes the remaining edges into linear motifs. We refer to this extraction order as ring $\rightarrow$ star $\rightarrow$ linear.
The rationale for this ordering is as follows. Ring motifs are extracted first because closed cycles are self-contained structures; removing them early leaves the remainder of the graph in a more tree-like form, which simplifies the subsequent steps and avoids their edges being fragmented across multiple linear blocks. Star motifs extraction is performed before linear motifs extraction because the linear motifs extraction algorithm acts as a catch-all step that consumes all remaining edges, leaving no edges for the star motifs extraction algorithm to process if the order were reversed. This cost difference becomes most apparent at high-degree vertices. If linear motifs extraction runs before star motifs extraction, a vertex of degree $2d$ is split across $d$ separate linear motifs rather than being grouped into a single star motif. As a result, the same vertex is counted $d$ times instead of once.

To validate that ring $\rightarrow$ star $\rightarrow$ line ordering is effective, as will become clear in the subsequent section, we enumerate all possible extraction orders on the minimum-edge representative of every LC equivalence class for graphs up to $n = 9$ nodes.  The ring $\rightarrow$ star $\rightarrow$ line ordering achieves the lowest mean resource overhead at every value of $n$ in this range, and no other orderings achieve a lower mean at any graph size. We therefore adopt this ordering throughout the remainder of this work. The following subsections describe each extraction stage in detail.





\begin{algorithm}[t]
\caption{Ring Motifs Extraction}
\label{alg:square_ring_extraction}
\begin{algorithmic}[1]
\State \textbf{Input:} Graph $G=(V,E)$; a set $\mathcal{T}$ of ring motifs
\State \textbf{Output:} Extracted ring motif collection $\mathcal{S}_{\mathrm{ring}}$ and residual graph $G$
\If{$E=\emptyset$}
    \State \Return $\emptyset$
\EndIf
\State Initialize $\mathcal{S}_{\mathrm{ring}} \leftarrow \emptyset$
\For{each $T \in \mathcal{T}$}
    \While{there exists a subgraph match of $T$ in $G$}
        \State Find a match $M$ of $T$ in $G$ using VF2
        \State Let $S_i$ be the matched motif with edge set $E(S_i)$
        \State Add $S_i$ to $\mathcal{S}_{\mathrm{ring}}$
        \State Remove $E(S_i)$ from $G$
    \EndWhile
\EndFor
\State Remove isolated vertices from $G$
\State \Return $\mathcal{S}_{\mathrm{ring}}, G$
\end{algorithmic}
\end{algorithm}

\paragraph{Stage 1: Ring Motifs Extraction}
Algorithm~\ref{alg:square_ring_extraction} extracts ring motifs from a graph $G=(V,E)$. To identify ring motifs, we adopt the VF2 subgraph isomorphism algorithm~\cite{cordella2004sub}.
Each ring motif is represented as a ring of $k$ nodes, and VF2 matches this motif against the target graph by mapping motif nodes to graph nodes one at a time. At each step, VF2 verifies that the required edges are preserved and that distinct motif nodes are mapped to distinct graph nodes. If either condition is violated, the search backtracks and tries a different assignment. A ring motif is successfully identified when all nodes and edges of the motif have been matched.

Using this matching procedure, we extract ring motifs by repeatedly identifying and removing them from the target graph. For each predefined ring motif of $k$ nodes, we search for all occurrences in the current graph and record the corresponding matched motif $S_i$. Once a match is found, $S_i$ is added to $\mathcal{S}_{\mathrm{ring}}$ and its edges $E(S_i)$ are removed from the graph to prevent reuse. This process repeats until no further matches can be found for any ring motif. After all ring motifs have been processed, isolated vertices are removed and the remaining graph is returned as the residual graph.

The time complexity of Algorithm~\ref{alg:square_ring_extraction} is dominated by the repeated VF2-based subgraph matching procedure. Since subgraph isomorphism is NP-complete, a single matching attempt is exponential in the worst case. For a ring motif $T\in\mathcal{T}$ with $|V(T)|=k$, the worst-case cost of a single matching attempt is $O(|V|^k)$. Since VF2 matching is applied repeatedly for each ring motif in $\mathcal{T}$ until no further match exists, the overall worst-case complexity is $O\!\left(\sum_{T\in\mathcal{T}} N_T\, |V|^{|V(T)|}\right)$, where $N_T$ denotes the number of matching attempts for ring motif $T$. In practice, however, the runtime is typically much lower because the ring motifs considered in our experiments are small and fixed in size, and matched edges are removed after each successful extraction, which progressively reduces the search space.



\begin{algorithm}[t]
\caption{Star Motifs Extraction}
\label{alg:star_extraction}
\begin{algorithmic}[1]
\State \textbf{Input:} Graph $G=(V,E)$
\State \textbf{Output:} Extracted star motif collection $\mathcal{S}_{\star}$ and residual graph $G$
\If{$|E| = 0$}
    \State \Return $\emptyset$
\EndIf
\State Compute vertex degrees $\deg_G(v)$ for all $v \in V$
\State Initialize $\mathcal{S}_{\star} \leftarrow \emptyset$
\State For each vertex $c \in V$, initialize an empty edge set $\mathcal{L}_c$
\For{each edge $(u,v) \in E$}
    \If{$\deg_G(u) > \deg_G(v)$}
        \State Assign center $c=u$, leaf $l=v$
    \ElsIf{$\deg_G(v) > \deg_G(u)$}    
        \State Assign center $c=v$, leaf $l=u$
    \Else
        \State Break tie by vertex index: $c=\min(u,v)$, $l=\max(u,v)$
    \EndIf
    \State $\mathcal{L}_c \leftarrow \mathcal{L}_c \cup \{(c,l)\}$
\EndFor
\For{each vertex $c \in V$}
    \If{$|\mathcal{L}_c| \ge 3$} \Comment{exclude degenerate stars (two-edge paths)}
        \State Let $S_i$ be the star motif with edge set $\mathcal{L}_c$
        \State $\mathcal{S}_{\star} \leftarrow \mathcal{S}_{\star} \cup \{S_i\}$
        \State $G \leftarrow G \setminus \mathcal{L}_c$
    \EndIf
\EndFor
\State Remove isolated vertices from $G$
\State \Return $\mathcal{S}_{\star}, G$
\end{algorithmic}
\end{algorithm}








\paragraph{Stage 2: Star Motifs Extraction}


After ring motifs extraction, Algorithm~\ref{alg:star_extraction} identifies star motifs (GHZ-type graph states).
For each edge $(u,v)$, we assign it to a candidate center based on vertex degree, directing the edge toward the higher-degree endpoint. When both endpoints have equal degree, we select the one with the smaller vertex index as the candidate center. This produces a grouping of edges around candidate centers. A group is retained as a star motif $S_i$ only if it contains at least three incident edges, which excludes degenerate cases that would otherwise reduce to linear motifs. The extracted edges $E(S_i)$ are then removed from $G$, and isolated vertices are discarded. This step systematically identifies star motifs and simplifies the remaining graph for the subsequent linear motif extraction stage.

The time complexity of Algorithm~\ref{alg:star_extraction} is $O(|V|+|E|)$. Computing vertex degrees requires $O(|V|+|E|)$ time. Each edge is then scanned once and assigned to a candidate center based on the degrees of its endpoints, which takes $O(|E|)$ time. The algorithm then traverses all vertices to examine the grouped edge sets and retain valid star motifs. Since each edge is assigned to exactly one candidate center and removed at most once, this traversal also takes $O(|V|+|E|)$ time. A final linear scan removes any remaining isolated vertices. The overall complexity is therefore $O(|V|+|E|)$.

\begin{algorithm}[t]
\caption{Linear \& Ring Motifs Extraction}
\label{alg:path_cleanup}
\begin{algorithmic}[1]
\State \textbf{Input:} Graph $G=(V,E)$
\State \textbf{Output:} Extracted linear motif collection $\mathcal{S}_{\mathrm{lin}}$ and ring motif collection $\mathcal{S}_{\mathrm{ring}}$
\State Initialize $\mathcal{S}_{\mathrm{lin}} \leftarrow \emptyset$, $\mathcal{S}_{\mathrm{ring}} \leftarrow \emptyset$
\If{$|E| = 0$}
    \State \Return $(\emptyset,\emptyset)$
\EndIf
    
\For{each connected component $H$ of $G$}
    \If{$|E(H)| = 0$}
        \State \textbf{continue}
    \EndIf
    
    \If{$H$ is 2-regular (i.e., $\deg(v)=2$ for all $v \in V(H)$)}
        \State Extract its simple cycle decomposition and add it to $\mathcal{S}_{\mathrm{ring}}$
        \State \textbf{continue}
    \EndIf
    
    \State Let $\mathcal{O}$ be the set of odd-degree vertices in $H$
    \State Partition $\mathcal{O}$ into arbitrary pairs, denoted by $\Pi$
    
    \State Let $H' \leftarrow H$ be a multigraph
    \For{each pair $(u,v)\in \Pi$}
        \State Add a \emph{virtual} edge $(u,v)$ to $H'$
    \EndFor
    
    \State Find an Eulerian circuit in $H'$
    \State Remove all virtual edges to obtain a collection of edge-disjoint trails $\mathcal{T}$
    \State Split each trail in $\mathcal{T}$ at repeated vertices until all motifs are vertex-simple
    
    \For{each motif $S_i$ in the post-processed collection}
        \If{$S_i$ is a ring motif}
            \State $\mathcal{S}_{\mathrm{ring}} \leftarrow \mathcal{S}_{\mathrm{ring}} \cup \{S_i\}$
        \Else
            \State $\mathcal{S}_{\mathrm{lin}} \leftarrow \mathcal{S}_{\mathrm{lin}} \cup \{S_i\}$
        \EndIf
    \EndFor
\EndFor
\State \Return $(\mathcal{S}_{\mathrm{lin}}, \mathcal{S}_{\mathrm{ring}})$
\end{algorithmic}
\end{algorithm}

\paragraph{Stage 3: Residual Linear Motifs Extraction}
The residual graph after ring and star motif extraction is typically sparse and locally path-like. Algorithm~\ref{alg:path_cleanup} decomposes this residual graph into linear and ring motifs.
For each connected component $H$, we first identify all odd-degree vertices and add virtual edges between them to obtain an augmented multigraph $H'$ in which every vertex has even degree. We then compute an Eulerian circuit on $H'$, and removing the virtual edges partitions the traversal into edge-disjoint paths. Each resulting path corresponds to a linear motif $S_i \in \mathcal{S}_{\mathrm{lin}}$, and any closed traversal corresponds to a residual ring motif $S_i \in \mathcal{S}_{\mathrm{ring}}$.
The time complexity of Algorithm~\ref{alg:path_cleanup} is $O(|V|+|E|)$ under a standard adjacency-list implementation. Connected component decomposition, degree inspection, odd-vertex pairing, and virtual edge insertion are all linear in the size of the input graph. The Eulerian circuit on each augmented component $H'$ can be found in linear time using Hierholzer's algorithm \cite{cormen2022introduction}. Removing the virtual edges and splitting the resulting trails into individual motifs can also be done in linear time by scanning each traversal once. The overall complexity is therefore $O(|V|+|E|)$.

Together, the three extraction stages produce a complete motif decomposition $\mathcal{S} = \mathcal{S}_{\mathrm{ring}} \cup \mathcal{S}_{\star} \cup \mathcal{S}_{\mathrm{lin}}$ of the target graph state $G$, where the ring motifs extracted in Stage 3 are merged with those from Stage 1 to form the complete $\mathcal{S}_{\mathrm{ring}}$. Each motif in $\mathcal{S}$ corresponds to an experimentally realizable photonic resource state.

\paragraph{From Decomposition to Synthesis}
The decomposition naturally leads to a constructive synthesis procedure. Each extracted motif can be prepared independently and then connected to others using Type-I fusion operations. Since the decomposition produces an edge-disjoint cover of the LC-equivalent graph state $G'$, every edge is accounted for exactly once. The LC-equivalent graph state $G'$ can therefore be reconstructed by fusing the extracted motifs, after which LC operations are applied to recover the target graph state $G$.

\section{Performance Evaluation}\label{evaluation}

\begin{figure*}[htbp]
    \centering
    \begin{subfigure}[b]{0.3\textwidth}
        \centering
        \includegraphics[width=\linewidth]{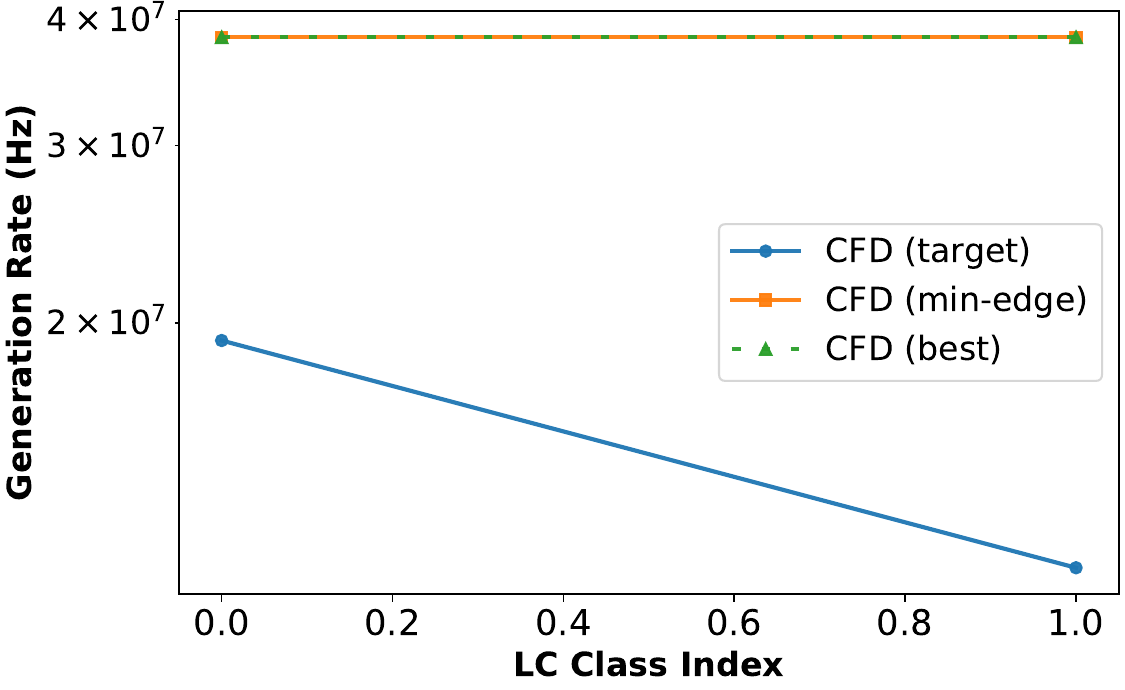}
        \caption{n=4}
    \end{subfigure}
    \hfill
    \begin{subfigure}[b]{0.3\textwidth}
        \centering
        \includegraphics[width=\linewidth]{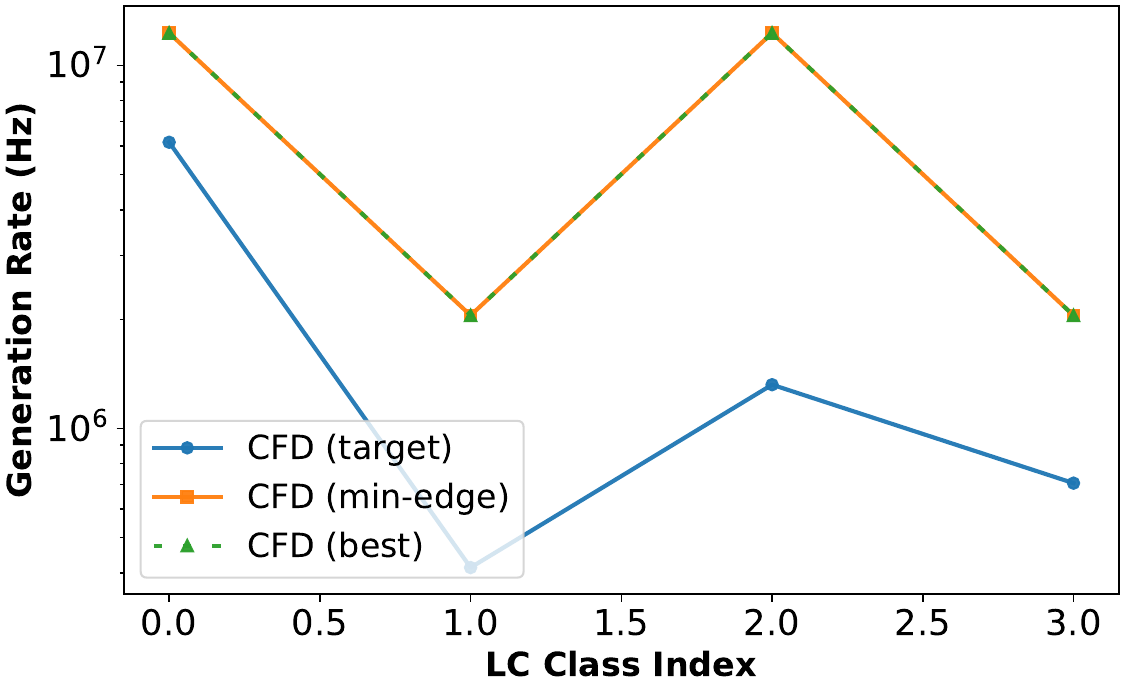}
        \caption{n=5}
    \end{subfigure}
    \hfill
    \begin{subfigure}[b]{0.3\textwidth}
        \centering
        \includegraphics[width=\linewidth]{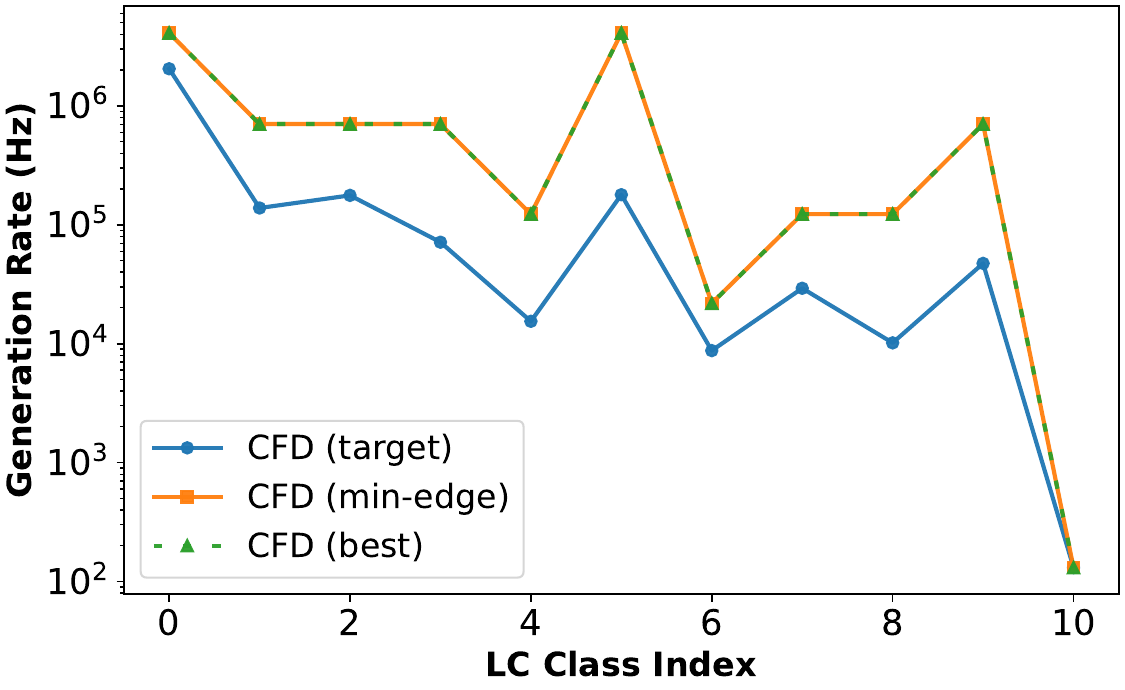}
        \caption{n=6}
    \end{subfigure}


    \begin{subfigure}[b]{0.3\textwidth}
        \centering
        \includegraphics[width=\linewidth]{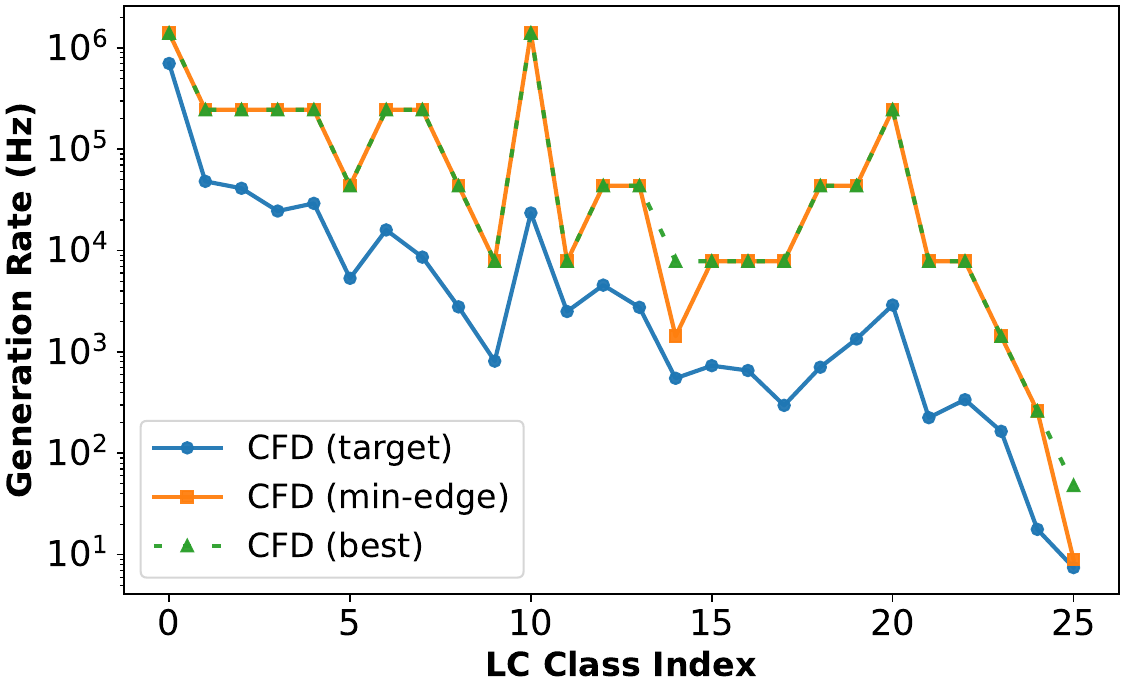}
        \caption{n=7}
    \end{subfigure}
    \hfill
    \begin{subfigure}[b]{0.3\textwidth}
        \centering
        \includegraphics[width=\linewidth]{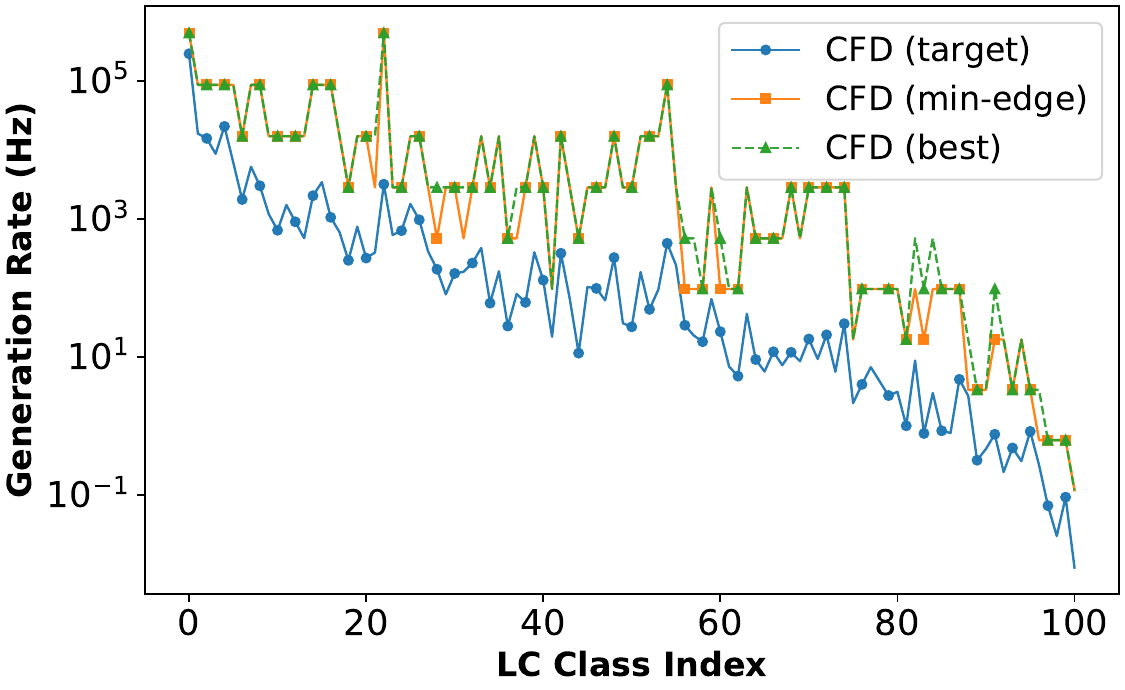}
        \caption{n=8}
    \end{subfigure}
    \hfill
    \begin{subfigure}[b]{0.3\textwidth}
        \centering
        \includegraphics[width=\linewidth]{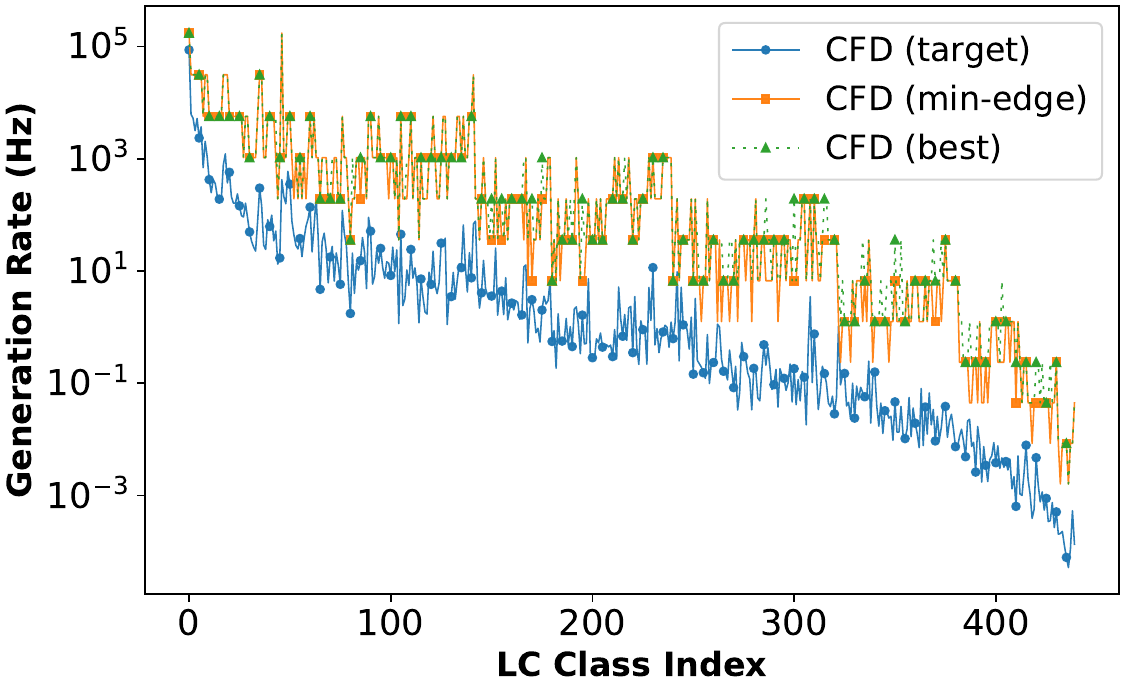}
        \caption{n=9}
    \end{subfigure}
    \caption{Photonic generation rates for graph states of size $n=4$ to $9$ across LC equivalence classes using the CFD framework. For each target graph state, CFD is applied to every graph state in its LC equivalence class and the generation rate is computed from the resulting motif decomposition. The plotted quantities correspond to the three cases defined in the Evaluation Method: CFD (min-edge), the generation rate obtained by applying CFD to the LC-equivalent graph state with the minimum number of edges; CFD (target), the generation rate obtained by applying CFD directly to the target graph state; and CFD (best), the minimum resource overhead achieved by CFD across the LC equivalence class.}


    \label{fig:generation_rate}
\end{figure*}




In this section, we evaluate the performance of the proposed CFD framework through numerical simulation using the Graph State Orbits data set \cite{adcock2020mapping}, synthesized 2D and 3D lattice graph states. We compare the resource overhead of CFD against existing methods for photonic graph state synthesis. 

\textbf{Graph State Dataset:} Part of our evaluations and comparisons are based on the Graph State Orbits Dataset, which includes all graph states for $n = 4$ to $9$ qubits \cite{adcock2020mapping}. In this dataset, for different sizes $n$, it has a different number of LC classes; in each class, it has all LC-equivalent graph states.

\textbf{Parameter Selection:} We use parameters that represent deterministic single-photon emission from InAs/GaAs quantum dots coupled to photonic cavities, with single-photon generation rate $R_1 \approx 6$ GHz and collection efficiency $\eta \approx 0.4$ \cite{somaschi2016near,singh2022optical}.

\textbf{Evaluation Method:} In our simulations, we mainly focus on graph states of size $n=4$ to $9$. We compare three cases: (i) CFD (min-edge), which applies CFD to the LC-equivalent graph state with the minimum number of edges; (ii) CFD (target), which applies CFD directly to the target graph state; and (iii) CFD (best), which reports the minimum resource overhead achieved by CFD across the LC equivalence class of the target graph state. The performance evaluation metrics include the generation rate $R(G)$ of the target graph state $G$ and the resource overhead, defined as the total number of photons consumed across all motifs:
\[
M = \sum_{i=1}^{k} c(S_i),
\]
where $c(S_i)$ denotes the photon cost of motif $S_i$. Specifically, $c(S_i) = |V(S_i)| + 1$ if $S_i$ is a ring motif, since generating a ring graph state of $|V(S_i)|$ photons requires preparing a linear graph state of $|V(S_i)| + 1$ photons and fusing its two end photons; and $c(S_i) = |V(S_i)|$ otherwise.

\subsection{Generation Rate Evaluation for Photonic Graph States}

We evaluate the photonic generation rate for graph states with $n=4$ to $9$ across LC equivalence classes using the proposed CFD framework. For each target graph state, we apply CFD to every graph state in its LC equivalence class and compute the corresponding generation rate based on the resulting decomposition. We report three quantities as defined in the Evaluation Method: CFD (min-edge), CFD (target), and CFD (best).

As shown in Fig.~\ref{fig:generation_rate}, for all sizes $n$, CFD (min-edge) consistently achieves higher generation rates than CFD (target). For small instances ($n=4$), the improvement remains modest, ranging from approximately $2\times$ to $3.4\times$. At $n=5$, the gap begins to widen, with improvements typically between $2\times$ and $9\times$. As the graph state size increases to $n=6$, the improvement becomes more pronounced, spanning from $1\times$ up to over $20\times$. For $n=7$, the gap further expands significantly, with improvements ranging from approximately $1.2\times$ to over $80\times$. At $n=8$, the improvement exhibits a wide variation, ranging from approximately $1.2\times$ to over $460\times$. For $n=9$, the improvement becomes consistently substantial, ranging from approximately $1.1\times$ to nearly $2400\times$. Overall, the advantage of CFD (min-edge) over CFD (target) grows rapidly with graph state size, increasing from a constant-factor improvement at small $n$ to one or even two orders of magnitude at larger scales.

We also observe that for $n \leq 6$, CFD (min-edge) exactly matches CFD (best). Although a gap between the two may appear in some LC equivalence classes as $n$ increases, they still coincide for most classes. This indicates that selecting the LC-equivalent graph state with the minimum number of edges is a highly effective strategy: in many cases, it matches the best generation rate observed within the LC equivalence class under CFD, and in most remaining cases it stays close to that best result.

\subsection{Comparison Across Graph State Synthesis Methods}

\begin{figure}
    \centering
    \includegraphics[width=0.8\linewidth]{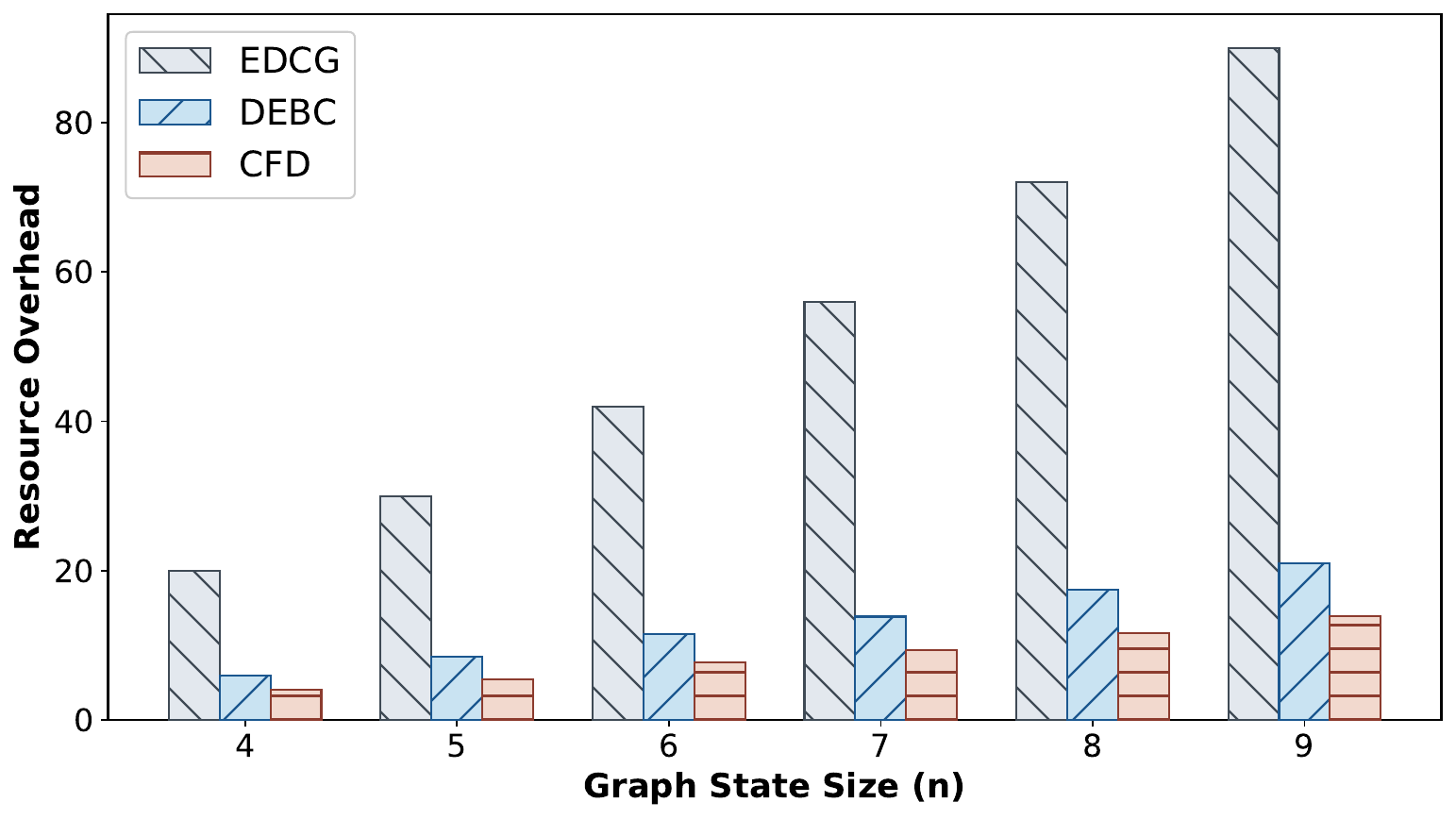}
    \caption{Resource overhead comparison of EDCG, DEBC, and the proposed CFD framework across graph states of size $n=4$ to $9$. EDCG incurs a fixed overhead of $n(n-1)$ regardless of target topology, while DEBC and CFD both take the minimum-edge LC-equivalent graph state as input. Lower resource overhead corresponds to higher photonic generation rate.}
    \label{fig:qubis_compariosn_baseline}
\end{figure}

We compare the proposed CFD method with two baselines: the Edge-Decorated Complete Graph (EDCG) method~\cite{meignant2019distributing} and Direct Edge-Based Construction (DEBC)~\cite{ji2024distributing}. 

EDCG provides a general way to generate an arbitrary graph state of size $n$. It first constructs a minimal tree on the $n$ vertices, and then applies star expansion to generate the required GHZ-type resource states. This process produces an edge-decorated graph, which can subsequently be projected onto the target graph state through fusion operations. In EDCG, the basic resource unit is the Einstein-Podolsky-Rosen (EPR) pair, a two-qubit maximally entangled state. For a target graph state of size $n$, the EPR cost of EDCG is $\frac{n(n-1)}{2}$, which corresponds to a resource overhead of $n(n-1)$.


DEBC, by contrast, directly uses EPR pairs to realize the target graph state. Its resource overhead is therefore $2|E|$, where $|E|$ denotes the number of edges in the graph state $G$. For both DEBC and CFD, we use as input the LC-equivalent graph state with the minimum number of edges, and then perform the corresponding decomposition on that graph.

As shown in Fig.~\ref{fig:qubis_compariosn_baseline}, CFD consistently outperforms both baselines in terms of resource overhead. 
For graph sizes from $n=4$ to $n=9$, CFD reduces the overhead by approximately 80.0\%, 81.7\%, 81.6\%, 83.4\%, 83.9\%, and 84.6\% relative to EDCG, respectively. 
Relative to DEBC, the corresponding reductions are approximately 33.3\%, 35.3\%, 32.5\%, 32.8\%, 33.2\%, and 34.0\%. 
This consistent gap shows that CFD can achieve a substantially more efficient construction than both EDCG and direct edge-based methods.

\begin{figure}
    \centering
    \includegraphics[width=\linewidth]{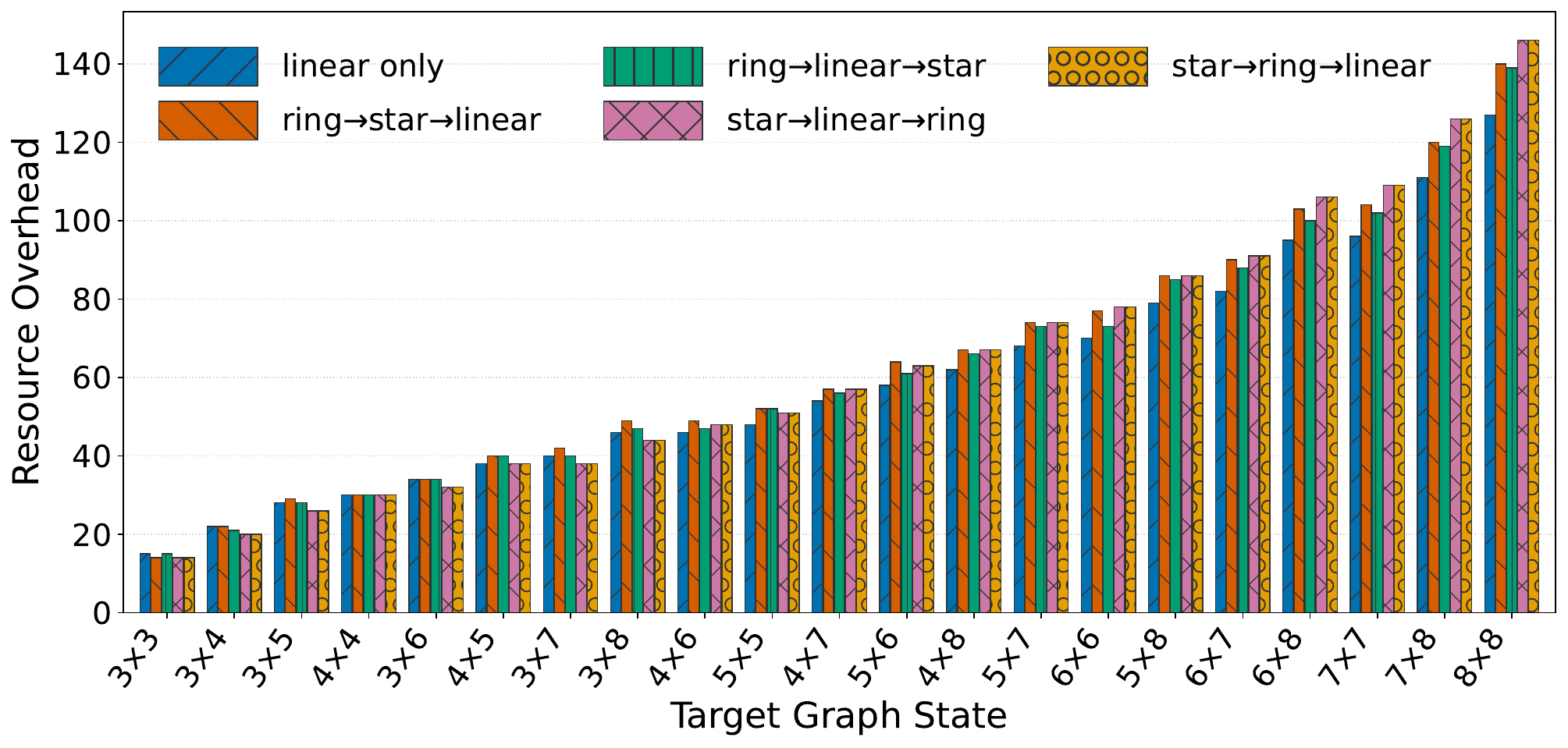}
    \caption{Resource overhead of different motif extraction policies for 2D cluster graph states of size $3 \times 3$ to $8 \times 8$. Each policy specifies a different ordering of ring, star, and linear motif extraction stages.}
    \label{fig:qubis_compariosn_2D}
\end{figure}

\begin{figure}
    \centering
    \includegraphics[width=\linewidth]{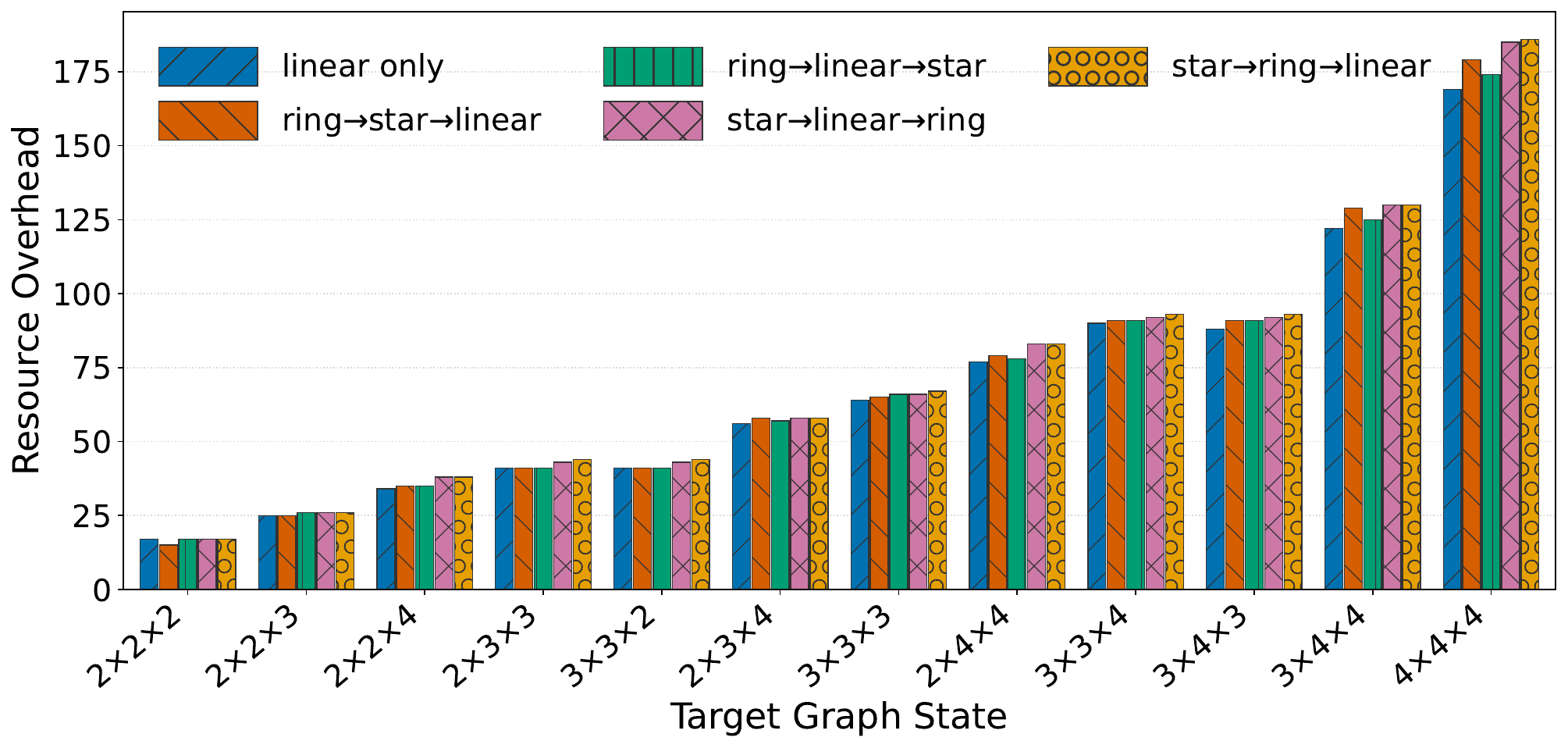}
    \caption{Resource overhead of different motif extraction policies for 3D cubic lattice graph states of varying sizes. Each policy specifies a different ordering of ring, star, and linear motif extraction stages.}
    \label{fig:qubis_compariosn_3D}
\end{figure}

\subsection{Resource-Overhead Analysis for 2D and 3D Lattice Graph States}

We further investigate which types of motifs are most effective for reducing the resource overhead of 2D and 3D lattice graph states. In this simulation study, we evaluate several motif extraction orders for decomposition. Specifically, we consider the following strategies:
(1) ring motif extraction, followed by star motif extraction, and then linear motif extraction;
(2) star motif extraction, followed by ring motif extraction, and then linear motif extraction;
(3) ring motif extraction, followed directly by linear motif extraction;
(4) star motif extraction, followed directly by linear motif extraction; and
(5) linear motif extraction only.
Here, ring motif extraction is performed using Algorithm~\ref{alg:square_ring_extraction}, star motif extraction uses Algorithm~\ref{alg:star_extraction}, and linear motif extraction uses Algorithm~\ref{alg:path_cleanup}, which additionally captures any residual ring motifs in the remaining graph.

Fig.~\ref{fig:qubis_compariosn_2D} presents the resource overhead of different motif extraction policies applied to 2D cluster states, with sizes ranging from $3 \times 3$ to $8 \times 8$. Overall, no single policy consistently achieves the lowest resource overhead across all instances. The relative performance of different policies varies with the size of the cluster. For smaller cluster states, policies that start with star motif extraction, such as \emph{star $\rightarrow$ ring $\rightarrow$ linear} and \emph{star $\rightarrow$ linear}, tend to perform better. This is possibly because the structure is relatively simple and can be efficiently represented by a few star motifs combined with short linear motifs. For larger cluster states, the \emph{linear only} policy consistently yields better results. This is because the grid structure becomes more regular as the size increases, and linear motif extraction can preserve this regularity without introducing additional overhead. In contrast, performing star motif extraction at an early stage may break the grid structure and lead to unnecessary resource consumption. Quantitatively, the \emph{linear only} policy reduces the resource overhead by up to $14\%$ compared to other policies for larger instances (e.g., $7 \times 7$ and $8 \times 8$).

Fig.~\ref{fig:qubis_compariosn_3D} shows the resource overhead of different motif extraction policies applied to 3D cubic lattice graph states. Similar to the 2D case, no single policy is optimal across all instances. However, the overall trend is more consistent. As the lattice size increases, the \emph{linear only} policy consistently achieves the lowest resource overhead. Compared to other motif extraction policies, it reduces the overhead by approximately $10\%$–$15\%$ for larger instances. This is because the 3D cubic lattice has a highly regular structure, and linear motif extraction preserves this regularity, while early star motif extraction tends to introduce unnecessary fragmentation and additional resource consumption. Overall, these results indicate that the most effective motif extraction policy depends on the target graph state structure, with highly regular lattice graph states favoring linear motif extraction, while irregular graph states benefit more from ring and star motif extraction.


\section{Conclusion}\label{conclusion}
In this work, we studied the problem of efficient photonic graph state synthesis from a cost-aware decomposition perspective. We proposed CFD, a structured three-stage framework that constructs target graph states from experimentally realizable motifs while minimizing fusion overhead and physical-qubit consumption.

Our results show that CFD significantly outperforms baseline constructions, achieving up to $84.6\%$ reduction in resource overhead across a range of graph state sizes. Due to the exponential dependence of generation rate on resource cost, these savings translate into substantial improvements in photonic generation rates, often spanning multiple orders of magnitude.

In addition, we observed that the decomposition cost varies significantly across the LC equivalence class of the same target graph state, revealing an important optimization opportunity that can be effectively exploited using simple structural heuristics. In particular, selecting the LC-equivalent graph state with the minimum number of edges provides a practical and highly effective strategy, matching the best generation rate observed within the LC equivalence class under CFD in many cases, and remaining close to it in most others.

Overall, our results demonstrate that combining structure-aware motif decomposition with LC equivalence enables efficient and scalable photonic graph state generation. We believe this perspective can inform the design of future synthesis algorithms and contribute to the development of practical photonic quantum computing architectures.

\balance

\bibliographystyle{IEEEtran}
\bibliography{info}

\begin{thebibliography}{10}
\providecommand{\url}[1]{#1}
\csname url@samestyle\endcsname
\providecommand{\newblock}{\relax}
\providecommand{\bibinfo}[2]{#2}
\providecommand{\BIBentrySTDinterwordspacing}{\spaceskip=0pt\relax}
\providecommand{\BIBentryALTinterwordstretchfactor}{4}
\providecommand{\BIBentryALTinterwordspacing}{\spaceskip=\fontdimen2\font plus
\BIBentryALTinterwordstretchfactor\fontdimen3\font minus \fontdimen4\font\relax}
\providecommand{\BIBforeignlanguage}[2]{{%
\expandafter\ifx\csname l@#1\endcsname\relax
\typeout{** WARNING: IEEEtran.bst: No hyphenation pattern has been}%
\typeout{** loaded for the language `#1'. Using the pattern for}%
\typeout{** the default language instead.}%
\else
\language=\csname l@#1\endcsname
\fi
#2}}
\providecommand{\BIBdecl}{\relax}
\BIBdecl

\bibitem{liu2025road}
J.~Liu, T.~Le, T.~Ji, R.~Yu, D.~Farfurnik, G.~Byrd, and D.~Stancil, ``The road to quantum internet: Progress in quantum network testbeds and major demonstrations,'' \emph{Progress in Quantum Electronics}, vol.~99, p. 100551, 2025.

\bibitem{hein2004multiparty}
M.~Hein, J.~Eisert, and H.~J. Briegel, ``Multiparty entanglement in graph states,'' \emph{Physical Review A—Atomic, Molecular, and Optical Physics}, vol.~69, no.~6, p. 062311, 2004.

\bibitem{raussendorf2003measurement}
R.~Raussendorf, D.~E. Browne, and H.~J. Briegel, ``Measurement-based quantum computation on cluster states,'' \emph{Physical review A}, vol.~68, no.~2, p. 022312, 2003.

\bibitem{browne2005resource}
D.~E. Browne and T.~Rudolph, ``Resource-efficient linear optical quantum computation,'' \emph{Physical Review Letters}, vol.~95, no.~1, p. 010501, 2005.

\bibitem{pant2019percolation}
M.~Pant, D.~Towsley, D.~Englund, and S.~Guha, ``Percolation thresholds for photonic quantum computing,'' \emph{Nature communications}, vol.~10, no.~1, p. 1070, 2019.

\bibitem{morley2018physical}
S.~Morley-Short, S.~Bartolucci, M.~Gimeno-Segovia, P.~Shadbolt, H.~Cable, and T.~Rudolph, ``Physical-depth architectural requirements for generating universal photonic cluster states,'' \emph{Quantum Science and Technology}, vol.~3, no.~1, p. 015005, 2018.

\bibitem{gimeno2015three}
M.~Gimeno-Segovia, P.~Shadbolt, D.~E. Browne, and T.~Rudolph, ``From three-photon ghz states to universal ballistic quantum computation,'' 2015.

\bibitem{gu2024fendi}
H.~Gu, Z.~Li, R.~Yu, X.~Wang, F.~Zhou, J.~Liu, and G.~Xue, ``Fendi: Toward high-fidelity entanglement distribution in the quantum internet,'' \emph{IEEE/ACM Transactions on Networking}, vol.~32, no.~6, pp. 5033--5048, 2024.

\bibitem{lindner2009proposal}
N.~H. Lindner and T.~Rudolph, ``Proposal for pulsed on-demand sources of photonic cluster state strings,'' \emph{Physical review letters}, vol. 103, no.~11, p. 113602, 2009.

\bibitem{cogan2023deterministic}
D.~Cogan, Z.-E. Su, O.~Kenneth, and D.~Gershoni, ``Deterministic generation of indistinguishable photons in a cluster state,'' \emph{Nature Photonics}, vol.~17, no.~4, pp. 324--329, 2023.

\bibitem{istrati2020sequential}
D.~Istrati, Y.~Pilnyak, J.~Loredo, C.~Ant{\'o}n, N.~Somaschi, P.~Hilaire, H.~Ollivier, M.~Esmann, L.~Cohen, L.~Vidro \emph{et~al.}, ``Sequential generation of linear cluster states from a single photon emitter,'' \emph{Nature communications}, vol.~11, no.~1, p. 5501, 2020.

\bibitem{de2024spin}
G.~de~Gliniasty, P.~Hilaire, P.-E. Emeriau, S.~C. Wein, A.~Salavrakos, and S.~Mansfield, ``A spin-optical quantum computing architecture,'' \emph{Quantum}, vol.~8, p. 1423, 2024.

\bibitem{rempe2024fusion}
P.~Thomas, L.~Ruscio, O.~Morin, and G.~Rempe, ``Fusion of deterministically generated photonic graph states,'' \emph{Nature}, vol. 629, no. 8012, pp. 567--572, 2024.

\bibitem{ding2016demand}
X.~Ding, Y.~He, Z.-C. Duan, N.~Gregersen, M.-C. Chen, S.~Unsleber, S.~Maier, C.~Schneider, M.~Kamp, S.~H{\"o}fling \emph{et~al.}, ``On-demand single photons with high extraction efficiency and near-unity indistinguishability from a resonantly driven quantum dot in a micropillar,'' \emph{Physical review letters}, vol. 116, no.~2, p. 020401, 2016.

\bibitem{schwartz2016deterministic}
I.~Schwartz, D.~Cogan, E.~R. Schmidgall, Y.~Don, L.~Gantz, O.~Kenneth, N.~H. Lindner, and D.~Gershoni, ``Deterministic generation of a cluster state of entangled photons,'' \emph{Science}, vol. 354, no. 6311, pp. 434--437, 2016.

\bibitem{thomas2022efficient}
P.~Thomas, L.~Ruscio, O.~Morin, and G.~Rempe, ``Efficient generation of entangled multiphoton graph states from a single atom,'' \emph{Nature}, vol. 608, no. 7924, pp. 677--681, 2022.

\bibitem{van2004graphical}
M.~Van~den Nest, J.~Dehaene, and B.~De~Moor, ``Graphical description of the action of local clifford transformations on graph states,'' \emph{Physical Review A}, vol.~69, no.~2, p. 022316, 2004.

\bibitem{van2004efficient}
------, ``Efficient algorithm to recognize the local clifford equivalence of graph states,'' \emph{Physical Review A—Atomic, Molecular, and Optical Physics}, vol.~70, no.~3, p. 034302, 2004.

\bibitem{bartolucci2023fusion}
S.~Bartolucci, P.~Birchall, H.~Bombin, H.~Cable, C.~Dawson, M.~Gimeno-Segovia, E.~Johnston, K.~Kieling, N.~Nickerson, M.~Pant \emph{et~al.}, ``Fusion-based quantum computation,'' \emph{Nature Communications}, vol.~14, no.~1, p. 912, 2023.

\bibitem{ji2024distributing}
T.~Ji, J.~Liu, and Z.~Zhang, ``Distributing arbitrary quantum graph states by graph transformation,'' \emph{arXiv preprint arXiv:2404.05537}, 2024.

\bibitem{cordella2004sub}
L.~P. Cordella, P.~Foggia, C.~Sansone, and M.~Vento, ``A (sub) graph isomorphism algorithm for matching large graphs,'' \emph{IEEE transactions on pattern analysis and machine intelligence}, vol.~26, no.~10, pp. 1367--1372, 2004.

\bibitem{cormen2022introduction}
T.~H. Cormen, C.~E. Leiserson, R.~L. Rivest, and C.~Stein, \emph{Introduction to Algorithms}, 4th~ed.\hskip 1em plus 0.5em minus 0.4em\relax MIT Press, 2022.

\bibitem{adcock2020mapping}
J.~C. Adcock, S.~Morley-Short, A.~Dahlberg, and J.~W. Silverstone, ``Mapping graph state orbits under local complementation,'' \emph{Quantum}, vol.~4, p. 305, 2020.

\bibitem{somaschi2016near}
N.~Somaschi, V.~Giesz, L.~De~Santis, J.~Loredo, M.~P. Almeida, G.~Hornecker, S.~L. Portalupi, T.~Grange, C.~Anton, J.~Demory \emph{et~al.}, ``Near-optimal single-photon sources in the solid state,'' \emph{Nature Photonics}, vol.~10, no.~5, pp. 340--345, 2016.

\bibitem{singh2022optical}
H.~Singh, D.~Farfurnik, Z.~Luo, A.~S. Bracker, S.~G. Carter, and E.~Waks, ``Optical transparency induced by a largely purcell enhanced quantum dot in a polarization-degenerate cavity,'' \emph{Nano letters}, vol.~22, no.~19, pp. 7959--7964, 2022.

\bibitem{meignant2019distributing}
C.~Meignant, D.~Markham, and F.~Grosshans, ``Distributing graph states over arbitrary quantum networks,'' \emph{Physical Review A}, vol. 100, no.~5, p. 052333, 2019.

\end{thebibliography}

\end{document}